\documentclass[icps3]{svjour}
\usepackage{epsf}
\begin{document}
%%%%%%%%%%%%%
\title{
Exciton-phonon droplets with Bose-Einstein condensate:
transport and optical properties   
}
\titlerunning{
Exciton-phonon droplets with Bose-Einstein condensate
}
%
% The running title should be in less than 70 characters.
%
\author{
D. Roubtsov\inst{1} 
\and 
Y. L{\'e}pine\inst{1}
\and 
I. Loutsenko\inst{2}
%%%%%%%%%
% etc
% \thanks{\emph{Present address:} Insert the address here if needed}%
%%%%%%%%%%%
}                     
%
% Insert author list for the running head here
%%%%%%%
\authorrunning{D. Roubtsov et al.}
%%%%%%%
% If the number of the authors is more than 3, only the first author
% should be listed and the others represented as et al.
%%%%%%%%
\institute{  
GCM et D{\'e}partement de physique, Universit{\'e} de Montr{\'e}al, 
C.P. 6128, succ. Centre-ville, Montreal, P.Q., H3C\,3J7, Canada
\and 
Lockheed Martin Canada et CRM, Universit{\'e} de Montr{\'e}al,
C.P. 6128, succ. Centre-ville, Montreal, P.Q., H3C\,3J7, Canada
}
\maketitle
\begin{abstract}
We discuss the possibility for a moving droplet of excitons and phonons to form a 
coherent state inside the packet. We describe such an inhomogeneous state in terms of  
Bose-Einstein condensation and prescribe it a macroscopic wave function.
Existence and, thus, coherency of such a Bose-core 
inside the droplet can be 
checked experimentally by letting two moving packets to interact.   
\end{abstract}
\section{Introduction}
\label{intro}
Nowadays, there is a lot of experimental evidence that  
excitons in semiconducting 
crystals  and heterostructures can form a strongly correlated state. 
In some cases, it can be assigned to the excitonic  Bose-Einstein condensate 
(BEC) 
\cite{a1}. 
As a rule, 
the excitons are prepared in the ground state with 
$\langle \hbar \,{\bf k} \rangle \simeq 0$ (or they are in a 
quasi-equilibrium state with some $T^{*}(t)$ 
cooling down toward the ground state). 
The conclusion about the presence of  
Bose-Einstein correlations among them 
is based on unusual properties of the direct PL signal 
from the excitonic cloud \cite{a1}. 

This article is motivated by experimental data on the 
transport properties of excitons 
in 3D crystals, such as  Cu$_{2}$0 
\cite{a2}, 
and  2D sheets in BiI$_{3}$ 
\cite{a3}. 
It is important to ``cook'' the  following initial conditions:   
a relatively dense cloud of excitons with the density of
$n_{\rm x} > n_{c}(T)$ (e.g., 
$n_{c}(T=2\,{\rm K})\approx 8.7 \times
10^{16}$\,cm$^{-3}$) 
to be  
in a moving state with \,
$\langle \hbar \,{\bf k}_{\rm x} \rangle \ne 0$.
In the case of Cu$_{2}$O crystals, these conditions can be achieved
because 
there is approximately the
same number of long wavelength acoustic phonons 
being formed in the same place as the excitons, $N_{\rm ph}\simeq
N_{\rm x}$ \, and \,$\varepsilon_{\rm ph} \simeq 1-5$\,{meV}.
Due to the exciton-phonon interaction and quasi-1D
geometry of the initial conditions,   
the acoustic phonons and excitons can form a packet with
$\langle \hbar \,{\bf k}_{\rm ph} \rangle \ne 0$ and 
$\langle \hbar \,{\bf k}_{\rm x} \rangle \ne 0$, so that 
\begin{equation}
\langle \hbar\, {\bf k}_{\rm x} \rangle 
\parallel 
\langle \hbar \,{\bf k}_{\rm ph} \rangle \parallel Ox
\end{equation}
and $\varepsilon_{\rm x} = m_{\rm x}v^{2}/2 
< m_{\rm x}c_{\rm s}^{2}/2 \sim
10^{-3}\,E_{\rm x}$. Here, $c_{\rm s}=4.5 \times 10^{5}$\,cm/s 
is the (longitudinal) speed of  sound, 
and $E_{\rm x}\simeq 0.15$\,eV is the exciton Rydberg, and 
$ m_{\rm x}\simeq 1.5\,m_{\rm e}$ is the exciton mass.

We assume that a Bose-correlated exciton-phonon core can be 
formed inside the
exciton-phonon packet under these conditions. 
Then, such a packet  
can move ballistically through the whole crystal at 
$T < T_{c}$, and the coherency of the Bose-core can be revealed 
by the packet-packet interaction or by stimulated scattering 
into the packet with the condensate. 
If $T>T_{c}$ or $n_{\rm x}< n_{c}(T)$, 
the exciton-phonon packet exhibits the standard diffusive behavior 
\cite{a2}.
Thus, the 3D droplets that supposed to contain 
the excitonic Bose-Einstein condensate are found in 
a spatially inhomogeneous  state
with the well-defined characteristic width $L_{\rm ch}$ 
in the direction of
motion. Note that this width $L_{\rm ch}=c_{\rm s}\,\tau_{\rm ch}$
can be different from the Bose correlation length of the Bose-core,
$L_{0}=c_{\rm s}\,\tau_{0}$. For example,  the
estimate 
$L_{0} \sim 10^{-1}\,L_{\rm ch}$, or 
\begin{equation}
\tau_{0} \simeq 50-60 \,{\rm ns}\, \sim \,  10^{-1}\,
\tau_{\rm ch} \simeq 0.2
- 0.4\,\mu{\rm s},
\end{equation}
can be extracted from the experiments on two packet interaction 
at $T<T_{c}$
\cite{a2}.  
The registered ballistic velocities of such excitonic packets 
turn out to be always less, but  relatively close to  
the longitudinal sound speed of the crystal, \,$v \,<\,c_{\rm s}$.
Note that the paraexcitons in pure Cu$_{2}$O crystals 
have an extremely large
lifetime,  $\tau \gg 13\,\mu$s, 
and a moving exciton with $\hbar k_{x} \sim m_{\rm x}\,c_{\rm s}$ cannot be 
converted into a photon directly.
Then, one 
can exclude the 
photons from simple models 
describing the transport of a {\it single} packet 
of excitons in a periodic medium. 
Moreover, we neglect the ortho-para exciton
conversion inside the formed packet of the {\it moving} paraexcitons,  
(i.e., $N_{\rm x,\, para}>N_{c}$, \,$N_{\rm x, \,ortho}\simeq 0$) 
in Cu$_2$O crystals \cite{KM}.
%%%%%%
% 
% This is not the case if... 
%
% Although transferring of the coherent excitonic field into 
% a coherent photon field is possible in theory 
% \cite{Fernandez},\cite{PS},
% the available experimental results are still disputable
% \cite{Lin},\cite{Butov},\cite{Snoke}.
%
%%%%%%

To understand  the physics of 
anomalous excitonic transport,  we assume  
that  the macroscopic wave 
function $\Psi_{0}\simeq \phi_{\rm o}\,e^{i\varphi_{\rm c}}$ can be
associated with the coherent part of the excitonic packet at
$T<T_{c}$.
Here, $\varphi_{\rm c}$ is the coherent phase of the condensate.
Indeed, the experimental results  \cite{a2},\cite{a3}
suggest the following decomposition
of the density of excitons in the packet,  
\begin{equation}
n({\bf x},t)= n_{\rm coh}({\bf x},t) + \Delta n({\bf x},t), 
\end{equation} 
where  $n_{\rm coh}({\bf x},t) \approx n_{\rm core}(x -vt)$
is the ballistic (superfluid) part of the packet, 
\begin{equation}
n_{\rm core}(x -vt) \simeq  \vert \Psi_{0}\vert^{2}(x-vt),  
\end{equation} 
and $ \Delta n({\bf x},t)$ is the noncondensed part of it.
The following decomposition can be written for the 
out-of-con\-den\-sa\-te part:
\begin{equation}
\Delta n({\bf x},t) = \langle \delta\hat{\psi}^{\dag} 
 \delta\hat{\psi}  ({\bf x},t) \rangle \approx 
\delta n_{\rm cloud}({\bf x},t)  + 
\delta n_{\rm tail}({\bf x},t).
\nonumber
\end{equation} 

%%%%%%%%%%%%%%%%%%%%%%
The challenging problem is how to describe 
the spatially inhomogeneous     
state of the droplet with the excitonic BEC inside 
in terms of $\Psi_{0}({\bf x},t)$ and
$\delta\hat{\psi}  ({\bf x},t)$, where 
$\delta\hat{\psi}$ is the ``fluctuating'' 
part of the exciton Bose-field. 
For example, within the quasistationary approximation, 
one has to calculate 
$n_{\rm core}({ x}/L_{0})$ and  
\begin{equation}
\Delta n({ x})\simeq  \delta n_{\rm o,\, cloud}({ x}/L_{\rm ch}) + 
\delta n_{\rm tail}
\end{equation}
and understand how 
the different characteristic lengthes and 
coherence properties appear in the theory with the Bose-condensate.
Note that if the excitonic packet moves  in a crystal (or another
semiconductor 
structure), it interacts with thermal pho\-nons, noncondensed excitons, 
impurities and other imperfections of the lattice, etc.. 
Then, the coherent core of the packet 
can be found in a quasi-stable 
state, and the fluctuations of $\phi_{\rm o}(x-vt)$ and, especially, 
$\varphi_{\rm c}(x,t)$ 
can be of a great importance for possible experimental 
verifications 
of their existence. 
 
\section{Exciton-Phonon Condensate}
\label{sec:1}

To obtain the necessary density of excitons $n_{\rm x}$ in the 
excitonic cloud and, thus,  meet the BEC
conditions, the crystals are irradiated  
by laser pulses with 
$\hbar \omega_{L} \gg E_{\rm gap}$, and the temperature of the
crystal is  $T\simeq 1\sim 5$\,K.
If  
the cross-section area $S$ of an excitation spot on the surface of the 
crystal 
can be made large enough, such as $S \approx  S_{\rm surf}$, 
the hot droplet of paraexcitons can 
acquire an average momentum  during its thermolization process 
($T^{*}(t) \rightarrow T$). 
Indeed,  
the phonon wind, 
or the flow of nonequilibrium phonons, blows unidirectionally from the surface 
into the bulk 
\cite{a4}
and transfers the nonzero momentum to the excitonic cloud, 
\begin{equation}
{\bf P}_{\rm x} \simeq N_{\rm x}\,\langle \hbar\, 
{\bf k}_{0}\rangle \ne 0 \,\,\,{\rm and}\,\,\,  
{\bf P}_{\rm x} \perp S_{\rm surf}, 
\nonumber
\end{equation}
see Fig. 1. 
As a result, the packet of moving excitons and nonequilibrium phonons 
of the phonon wind ($N_{\rm ph}\simeq N_{\rm x}$)
is actually the system that undergoes the transition toward developing  
the Bose-Einstein correlations at $T^{*}<T_{c}$.

%%%%%%%%%%%%%%%%%%%%%%%%%%%%
\begin{figure}
\begin{center}
\leavevmode
\epsfxsize = 200pt
\epsfysize = 100pt
\epsfbox{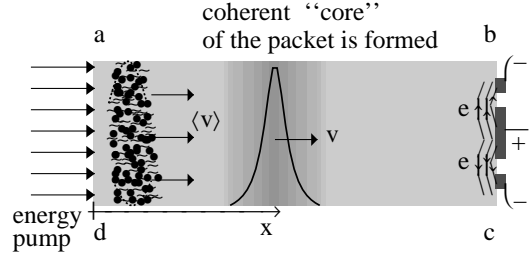}
\end{center}
%%%%%%%%%%%%%%%
\caption{
 A medium, in which the exciton-phonon droplet can propagate, is presented 
 in the form of  the channel `abcd' on this Figure. 
%
% It has the dimensions $\vert {\rm ab}\vert=L$, \,
% $\vert {\rm bc}\vert \simeq \sqrt{S_{\perp}}$,
% and $L\gg L_{0}$, where  $L_{0}$ is the characteristic width of the soliton.
%
After some amount of energy has been  pumped into the medium 
during a short time interval and absorbed near a boundary,
a localized excited state is formed near the face `ad'.
%
% It is schematically shown on Figure 1\,(a) as a mixture of excitons and phonons.  
%
If there is a 
mechanism of the momentum transfer to the excited state, 
the droplet begins to move toward
the opposite face `bc' with the velocity $\langle v\rangle$. 
Then, such conditions can favor the appearance of an inhomogeneous coherent 
state inside the droplet if the average density of the excitons $n_{\rm x}>n_{c}(T)$. 
%
% moving ballistically along the axis $Ox$.
% In other words, a sort of Bose-condensate appears because of the effective attraction
% among the bosons (excitons) at $T<T_{c}$, see Fig. 1\,(c). 
%
The profile of the excitonic part of it,
$n_{\rm core}(x,t) \simeq \vert \Psi_{\rm o}(x,t) \vert^{2}$, 
is shown by the bold line  and the intensity of the elastic (phonon) part,
$\partial_{x}u_{{\rm o},\,x}(x,t)$, is represented  
by changements of the intensity of the background color.
When the packet reaches the face `bc', the total density of excitons $n(x)$ 
is converted into an electric current, $i(t)$. 
}
\end{figure}
%%%%%%%%%%%%%%%%%%%%%%%%%%%%
\begin{figure}[b]
\begin{center}
\leavevmode
\epsfxsize=215pt % will enlarge or reduce the postscript figures based on the xsize
\epsfysize=200pt
\epsfbox{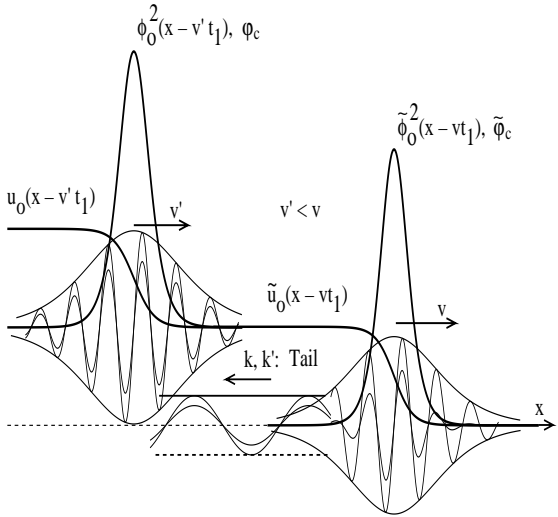} % postscript image file name
\end{center}
\caption
{ Two ballistic packets with two exciton-phonon condensates
 (\,${\rm e}^{i\varphi_{c}(x,t)}\phi_{\rm o}(x-vt)
 \cdot u_{\rm o}(x-vt)\,\delta_{1j}$\,) inside
 were created with the same concentration of excitons 
 and the same ballistic velocity, $v$, in a crystal. 
% (The face `ad' on Fig. 1 
% was irradiated by the same laser pulse twice.)  
 The time delay between them is 
 a free parameter.
 Then, two different interaction regimes are possible. The first one 
 corresponds to the case in which the Bose-cores of the packets
 overlap. This is a strong interaction case, and the packets
 can merge into one droplet.
 The second regime, in which the Bose-cores do not overlap,
 is the case of weak interaction between the packets. 
 It is depicted on this Figure. 
 However, the second moving packet (the left one)  
 can ``feel'' the first packet (the right one)
%% ${\rm e}^{i\tilde{\varphi}_{c}(x,t)}\tilde{\phi}_{\rm o}(x-vt)
%% \cdot \tilde{u}_{\rm o}(x-vt)\delta_{1j}$,
 through the    
 interaction with the exciton-phonon tail of the first one. 
 As a result, the second packet slows down, $v'<v$,  
 and becomes more broad.    
 }
\label{fig_two_solitons}
\end{figure}

Let us assume that 
the condensate has been already formed inside the moving excitonic droplet, and 
the following representation of the exciton Bose-field holds:  
$\hat{\psi}=\Psi_{0} + \delta\hat{\psi}$.
%
%% (As the kinetics of condensate formation is not a subject 
%% of this Letter, we assume  $T_{\rm cloud} \simeq T$ and 
%  $T\rightarrow 0$.)
%
%    
For the  displacement field of the crystal $\hat{\bf u}$,  
we introduce 
a nontrivial coherent part too, i.e., $\hat{\bf u}={\bf u}_{0} + 
\delta\hat{\bf u}$, 
\,${\bf u}_{0} \ne 0$.
The important property of the exciton-phonon condensate is a kind of
the
self-consistency condition, roughly, 
\begin{equation}
\partial {\bf u}_{0}(x-vt) \propto \vert \Psi_{0}
(x-vt)\vert^{2}. 
\end{equation}
In these terms,  
the moving packet contains both 
the macroscopically occupied exciton-phonon condensate, 
or the Bose-core
$\Psi_{0}({\bf x},t) \cdot {\bf u}_{0}({\bf x},t)$,
and out-of-condensate excitons and phonons.
The macroscopic wave function of excitons $\Psi_{0}(x,t)$ 
is normalized as follows: 
\begin{equation} 
\int\!\!\vert\Psi_{0}\vert^{2}(x,t)\,d{\bf x}
=S\! \int\!\!\phi_{\rm o}^{2}(x/L_{0})\,dx = N_{\rm o} \gg 1,
\label{norma} 
\end{equation}
where $N_{\rm o}$ is the macroscopic number of condensed excitons, and,
generally, 
\,$N_{\rm o}(T) < N_{\rm x}$. Within the quasistationary 
approximation, we have \,$N_{\rm o}(T)={\rm const}$ and 
$\delta N(T)={\rm const}$
\,that implies 
(quasi)stability of the moving packet with the Bose-core 
during the finite observation time.

To model the ballistic motion of a single packet, we use  
the following ansatz for the Bose-core of the packet:  
\begin{equation}
\Psi_{0}(x,t) =
{\rm e}^{-i (\tilde{E}_{g} + m_{\rm x}v^{2}/2 -\vert{\mu}\vert )\,t}
{\rm e}^{ i(\varphi_{\rm c} + k_{0}x)}\,
\phi_{\rm o}(x-vt),
\label{ansatz1}
\end{equation}
%%%%%%%%%%%%%%%%%%%%%%%
\begin{equation}
 u_{0\,j}(x,t)=u_{\rm o}(x-vt)\,\delta_{1j},
\label{ansatz2}
\end{equation}
where \,$\tilde{E}_{g}= E_{\rm gap}-E_{\rm x}$, 
% $E_{\rm x}$ is the exciton Rydberg,  
$\varphi_{\rm c}={\rm const}$ is  the macroscopic phase,
\,$\hbar k_{0}=m_{\rm x} v$, 
and $\mu = \mu (N_{\rm o})<0$ is 
the effective chemical potential of the condensate. 
At $T\ll T_{c}$, one can disregard the interaction between the 
Bose-core and the out-of-condensate cloud and write down the following 
equations on the envelope functions $\phi_{\rm o}(x/L_{0})$ and 
$u_{\rm
o}(x/L_{0})$ \cite{a5}:
%%%%%%%%%%%%%%%%%%%%%%%%%%%
%% \begin{equation} 
$$
-\vert{\mu}\vert\,\phi_{\rm o}(x)= -(\hbar^{2}/2m_{\rm x}) 
\partial_{x}^{2}\phi_{\rm o}(x)-
\vert \tilde{\nu}_{0}\vert \,\phi_{\rm o}^{3}(x)
 +
\tilde{\nu}_{1}\,\phi_{\rm o}^{5}(x), 
$$
%% \label{1Deq}
%% \end{equation}
%%%%%%%%%%%%%%%%%%%%%%%%%%%%%%%
\begin{equation} 
\partial_{x}u_{\rm o}(x) \approx -{\rm const}_{0}\,\phi_{\rm o}^{2}(x) +
{\rm const}_{1}\, \phi_{\rm o}^{4}(x).
\label{1Deq1} 
\end{equation}
%%%%%%%%%%%%%%%%%%%%%%%%%%%%
At \,$T\ne 0$, \,$T<T_{c}$, we choose the quasistationary approximation
to write out the decomposition of  
the exciton and phonon 
fields of the moving droplet, 
$$
\hat{\psi}_{0}({\bf x},t)=
\exp\bigl(i\varphi_{c }(x,t)\,)
\bigl\{\phi_{\rm o}(x-vt) + 
\delta\hat{\psi}_{\rm o}(x-vt,\,{\bf x}_{\perp},\,t)
\bigr\},
$$
\begin{equation}
\hat{u}_{{0},\,j}({\bf x},t)= u_{\rm o}(x-vt)\,\delta_{1,j}
+\delta\hat{u}_{{\rm o},\,j}(x-vt,\,{\bf x}_{\perp},\,t).
\end{equation}
Then, the following correlation functions have to be included into
an analog of Eq. (\ref{1Deq1}):   
the `anomalous' ones, such as $\tilde{\rm m}({ x})=\langle
\delta\hat{\psi}_{\rm o}\, \delta\hat{\psi}_{\rm o}
\rangle $, 
%%%%%%%%%%%%%%%%%%
%%%%%%%%%%%%%%%%%%
the exciton-phonon correlators, such as $
\tilde{q}_{j}= 
\langle \partial_{j}\delta\hat{u}_{{\rm o},\,j}\,
\delta\hat{\psi}_{\rm o}(x,\,{\bf x}_{\perp},\,t) \rangle $,
%%%%%%%
%%%%%%%%%%%%%%%%%%%
and the out-of-condensate density of the excitons and phonons,
$$
\delta n_{\rm o}(x)=
\langle \delta\hat{\psi}_{\rm o}^{\dag}
\delta\hat{\psi}_{\rm o}(x,\,{\bf x}_{\perp},\,t)
\rangle  
\,\,
{\rm and }
\,\,
Q_{xx}(x)=\langle
(\partial_{x}\delta\hat{u}_{{\rm o},\,x})^{2}\rangle.
$$ 
It is possible to generalize Eq. (\ref{1Deq1}) 
to the case of $T\ne 0$. 
Here, we write it in the following (tractable) form: 
$$ 
-(\,\vert{\mu}\vert+ \delta\mu\,)\,\phi_{\rm o}(x)=
 -(\hbar^{2}/2m^{*}) 
\partial_{x}^{2}\phi_{\rm o}(x) + 
(\tilde{\nu}_{0}+ \delta{\nu}_{0})\,\phi_{\rm o}^{3}(x)
 $$
\begin{equation} 
+\,(\tilde{\nu}_{1}+\delta{\nu}_{1})\,\phi_{\rm o}^{5}(x),
  \label{1DeqT}
\end{equation}
%%%
\begin{equation} 
\partial_{x}u_{\rm o}(x) \approx -{\rm const}_{0}'\,\phi_{\rm o}^{2}(x)+ 
{\rm const}_{1}'\,\phi_{\rm o}^{4}(x)  
-\vert{\rm  const}_{\rm tail}\vert .
%\label{1Deq1T} 
\nonumber
\end{equation}
To a first approximation, 
we disregard all the dissipation terms and assume $T^{*}\simeq T$. 
Then, the corrections to the effective chemical
potential $\vert\mu\vert$,
the x-x and x-ph interaction vertices, $\tilde{\nu}_{j}$ and 
${\rm const}_{j}$, depend on the temperature $T$ through the
above mentioned correlation functions. 
This means Eq. (\ref{1DeqT})
has to be solved together with the equations on the
out-of-condensate
excitons and phonons,  
$\delta\hat{\psi}_{\rm o}$, $\delta\hat{\psi}_{\rm o}^{\dag}$, and 
$\delta\hat{u}_{{\rm o},\,j}$. 
Note that at $T\ne 0$, $T<T_{c}$ the effective interaction 
vertices in Eq. (\ref{1DeqT}) are strongly renormalized in comparison
with the ``bare'' ones staying in the Hamiltonian of the
exciton-phonon system. For example, the localized solution for 
$\phi_{\rm o}(x/L_{0})$ (a kind of the ``bright'' soliton of the 
cubic-quintic
NLS equation) exists due to the effect of such a strong renormalization,
$\tilde{\nu}_{0} + \delta{\nu}_{0}<0$ and 
$\tilde{\nu}_{1}+ \delta{\nu}_{1}>0$.

Note that the dependence of $n_{\rm o}(x/L_{\rm o})$ on 
$N_{\rm o}$ is highly nonlinear.  As a rough estimate, 
we can write $n_{\rm o} \propto N_{\rm o}^{2}$ ($T\rightarrow 0$).
This suggests that the estimates presented in \cite{ST_PRL}  
have to be revisited.
Some qualitative results obtained for the coherent exciton-phonon
packets
are presented on Fig. 2.

Finally, we speculate on the possibility to convert the coherent
exciton field into the coherent photon one 
by colliding two moving exciton-phonon packets.
Such an experiment could reveal how rigid is the
macroscopic phase of the Bose-core (if any).
If one can prescribe a 
macroscopic phase $\varphi_{c}$ to each Bose-core of the   
moving packets, the face-to-face collision of them  could result 
in the many-photon production. 
For example, the ``jets'' of  phonons originated from the condensate
interaction (the core-to-core collision) seem to be   
highly directional in space and with a low noise level, see Fig. 3.

\begin{figure}[b]
\begin{center}
\leavevmode
\epsfxsize = 200pt
\epsfysize = 180pt
\epsfbox{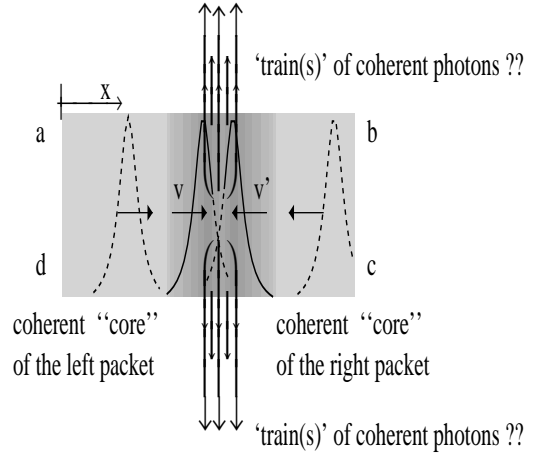}
\end{center}
%%%%%%%%%%%%%%%
\caption{
After some amount of energy 
($\delta E_{\rm left}= \delta E_{\rm right}$) 
has been  pumped into the medium 
during a short time interval $\delta t$ and 
absorbed near the left and right boundaries,
two localized excited states are formed near the faces `ad' and `bc'.
If there is a 
mechanism of the momentum transfer to the excited state, 
the droplets begin to move toward
the opposite faces with 
the velocities 
$\langle v\rangle \approx \langle v'\rangle$. 
Such conditions can favor the appearance of a coherent boson-phonon
state (an analog of Davydov soliton) 
inside both the left and right droplets.
The collision of such exciton-phonon droplets seems to be analogous to 
the heavy nucleus collisions that result in the production of
pion jets. In Cu$_{2}$O crystal, one can expect the production of 
two photon ``jets'', and coherence of the phonons in such jets 
can be checked experimentally.
}
\end{figure}

\end{document}